\documentclass[12pt]{iopart}

\usepackage{graphics,epsfig,subfig}
\usepackage{amssymb,amsfonts,bm}
\usepackage{mathrsfs}
\usepackage{color,psfrag}
\usepackage{iopams}
\usepackage{hhline}

\newcommand{\nn}{\nonumber}
\newcommand{\bb}{\begin{eqnarray}}
\newcommand{\ee}{\end{eqnarray}}

\renewcommand{\eref}[1]{(\ref{#1})}
\renewcommand{\Re}{{\rm Re}}
\renewcommand{\Im}{{\rm Im}}

\newcommand{\diag}{\mathrm{diag}}

\renewcommand{\fl}{\hspace{-1cm}}
\newcommand{\com}{\textcolor{black}}

\newcommand{\vk}{v}
\newcommand{\tvk}{\tilde\vk}
\newcommand{\bvk}{\bm{v}}
\newcommand{\nk}{n}
\newcommand{\ak}{K}

\newcommand{\bic}{\bm{j}}
\newcommand{\mass}{M_0}
\newcommand{\xl}{z}

\newcommand{\veps}{w}
\newcommand{\vl}{{\vk_1^*}}
\newcommand{\damp}{\gamma}
\newcommand{\tveps}{\tilde\veps}
\newcommand{\amp}{u}
\newcommand{\aamp}{a}
\newcommand{\func}{{f}}
\newcommand{\aq}{\alpha}
\newcommand{\wk}{w}

\begin{document}
\title[Origin of the spontaneous oscillations]{Origin of the spontaneous oscillations in
a simplified coagulation-fragmentation system driven by a source}
\author{Jean-Yves Fortin}
\address
{\com{Laboratoire de Physique et Chimie Th\'eoriques,
CNRS UMR 7019, 
\\ Universit\'e de Lorraine,
F-54000 Nancy, France}
}
\ead{jean-yves.fortin@univ-lorraine.fr}
\begin{abstract}{We consider a system of aggregated clusters of particles, subjected to coagulation and fragmentation processes with mass dependent rates. Each monomer particle
can aggregate with larger clusters, and each cluster can fragment into individual monomers with a rate
directly proportional to the aggregation rate. The dynamics of the cluster densities is governed by a set of Smoluchowski equations, and we consider the addition of a source of monomers at constant rate. The whole dynamics can be reduced to solving a unique non-linear differential equation which displays self-oscillations in a specific range of parameters, and for a number of distinct clusters in the system large enough. This collective phenomenon is due to the presence of a fluctuating damping coefficient and is closely related to the Li\'enard self-oscillation mechanism observed in a more general class of physical systems such as the van der Pol oscillator.}
\end{abstract} 


\section{Introduction}

The processes of aggregation and fragmentation are found in a large variety of systems at different
levels of scale. We can mention various physical phenomena that fit into this class of problems, such as foam dynamics with coalescence of soap bubbles, the stability of atmospheric particles of different sizes which aggregate due to the van der Waals forces, or even internet large scale networks where nodes can acquire additional links according to their attachment preference weight \cite{book:Krapivsky}. On larger scales, it is known that aggregation plays an important role in the formation of planetary rings \cite{Cuzzi:2010} or dust clouds induced by gravitational forces, and in the formation of galaxy clusters \cite{Brilliantov:2015,Brilliantov:2018a}. In contrast, fragmentation acts as a counterweight to the aggregation phenomenon, leading in many cases to a stationary state for the mass or size distribution of the particle clusters. The mechanisms of fragmentation 
have diverse origins. For example, the shattering of heavy particles that collide leads to smaller particles, or tidal forces between large masses can break them apart because of the inhomogeneity of the gravitation between different parts of the same body. Thermal fluctuations are also responsible for the desegregation of large polymers in solutions.

A general kinetic theory that incorporates both the aggregation and fragmentation is usually constructed from a set of Smoluchowski type equations that govern the dynamics of the mass densities of each individual object. In the long time limit, the steady state distribution exhibits in many models, according to the type of kernel chosen for the aggregation or fragmentation processes, a power law behavior associated with an exponential decay \cite{White:1981,Ziff:1985,DaCosta:2015,Krapivsky:2017,Bodrova:2019}. A summary of several analytical solutions for the density distribution is presented in Table \ref{tab:table1}, for different kinds of kernels and eventually in the presence of a particle source.
Apart from these time independent steady states, leading to an equilibrium distribution of masses, one can also observe collective oscillations in a narrow window of parameter range \cite{Ball:2012,Connaughton:2018,Brilliantov:2018}, see Table \ref{tab:table2}. However it is not yet clear how these oscillations arise from the complex dynamics involving different clusters of particles, since it is a highly non-linear process and
the Smoluchowski equations are usually quadratic in the densities and exact solutions can not be found in general, except for small sets of equations \cite{Buchstaber:2012a,Calogero:2020}, and this is an ongoing challenge in the field of mathematics.

In this paper we would like to investigate analytically the nature and origin of these oscillations in a simplified model of aggregation and fragmentation process. Here we consider a set of $N\gg 1$ possible distinct masses $M_k=kM_1$, $k\le N$, with their density $n_k$, and we restrict in this manuscript the dynamics to the aggregation process between the monomer particles and bigger masses $M_1+M_k\rightarrow M_{k+1}$, and the fragmentation of large masses into the smallest units $M_1+M_k\rightarrow (k+1)M_1$, see figure \ref{fig1_model}. We also add a source of monomers $M_1$ at constant rate and eventually a sink for the largest clusters $M_N$, which conserves the total mass of the system.

\begin{table}[tbp]
\caption{Asymptotic distributions for several models of aggregation-fragmentation in the steady state. The notations are the following: Aggregation kernel $K_{ij}$, fragmentation kernel $F_{ij}$, spontaneous fragmentation rate $F_k$, and source of $k$-mass clusters $I_k$. $\theta=\nu-\mu\ge 0$ and $\beta=\mu+\nu$. The mass density is normalized to unity.}
\label{tab:table1}
{\renewcommand{\arraystretch}{1.2}
\begin{tabular}{c|c|c}
Kernels $K_{ij}$, $F_{ij}$, $F_k$, and current $I_k$ & Asymptotic distribution & References \\
\hhline{===}
\footnotesize
$K_{ij}=1$, constant kernel, $F_{ij}=0$
&\footnotesize
$n_k(t)\simeq 4t^{-2}
e^{-2k/t}$ & \cite{Wattis:2006} \\
\hline
\footnotesize
$K_{ij}=(i+j)/2$, additive kernel, $F_{ij}=0$
&\footnotesize
$n_k(t)\simeq e^{-2t}g(ke^{-t})$, $g(z)=e^{-z/2}(2\pi z^3)^{-1/2}$ & \cite{Wattis:2006} 
\\
\hline
\footnotesize
$K_{ij}=ij$, multiplicative kernel, $F_{ij}=0$
&\footnotesize
$n_k(t)\simeq (2\pi)^{-1/2}k^{-5/2}t^{-1}$ & \cite{Wattis:2006} \\
\hline
\footnotesize $K_{ij}=ij$, $F_{ij}=0$, $F_k=\nu k$, $\nu\ge 2$ (gelation for $\nu\le 2$) & \footnotesize $n_k\simeq
\frac{(\nu-2)k+(1+\alpha)}{2\sqrt{\pi}}e^{-k\log(\nu/2)}k^{-7/2}$
& \cite{Vigil:1988} \\
\hline
\footnotesize
$K_{ij}=i^{\mu}j^{\nu}+i^{\nu}j^{\mu}$, $F_{ij}=0$, \footnotesize$I_k=I\delta_{k1}$, $\theta<1$ and $\beta<1$ & \footnotesize
$n_k\simeq \left (\frac{I(1-\theta^2)\cos(\pi\theta/2)}{4\pi}\right )^{1/2}
k^{-(3+\beta)/2}$ & \cite{Hayakawa:1987,Hayakawa:1988}
\\
\hline
\footnotesize
$K_{ij}=i^{\mu}j^{\nu}+i^{\nu}j^{\mu}$, $F_{ij}=\lambda K_{ij}$, $\mu=\nu=\beta/2$
&\footnotesize
$n_k\simeq n_1\frac{(1+\lambda)^2}{\sqrt{\pi}(1+2\lambda)} 
e^{-k}k^{-(3+\beta)/2}$ & \cite{Connaughton:2018} \\
\hline
\footnotesize
$K_{ij}=1$, $F_{ij}=\lambda$, $F_k=\nu$
&\footnotesize
$n_k\simeq c(4\pi)^{-1/2}e^{-\nu k/2}k^{-3/2}$, &
\\
& \footnotesize $c=(\lambda-\nu+\eta)/(1+2\lambda)$ &
\\
& \footnotesize $\eta=\sqrt{(\lambda-\nu)^2+2\nu(1+2\lambda)}$ &
\cite{Bodrova:2019}
\\
\hline
\footnotesize
$K_{ij}=(ij)^{\mu}$, $F_{ij}=\lambda$, $F_k=0$
&\footnotesize
$n_k\simeq \frac{\lambda^{1-2\mu}}{\Gamma(1/2-\mu)}e^{-\lambda^2 k}k^{-3/2-\mu}$ &
\cite{Bodrova:2019}
\\
\hline
\footnotesize $K_{1i}=K_{i1}=1$, $F_{1i}=F_{i1}=\lambda$, $\lambda>\lambda_c$ & \footnotesize $n_k
\simeq \frac{\lambda}{1+\lambda}e^{-k\log(1+\lambda)}$
&\cite{Krapivsky:2017}
\\
\hline
\footnotesize $K_{1i}=K_{i1}=i$, $F_{1i}=F_{i1}=\lambda i$, $\lambda>\lambda_c$ & \footnotesize$n_k
\simeq \lambda e^{-k\log(1+\lambda)}k^{-1}$
&\cite{Krapivsky:2017}
\\
\hline
\end{tabular}}
\end{table}
%
%
\begin{table}[tbp]
\caption{Models with stable oscillations}
\label{tab:table2}
{\renewcommand{\arraystretch}{1.2}
\begin{tabular}{c|c|c}
Kernels $K_{ij}$, $F_{ij}$, and current $I_k$ & Parameter range for stable oscillations & References \\
\hhline{===}
\footnotesize
$K_{ij}=i^{\mu}j^{\nu}+i^{\nu}j^{\mu}$, $I_k=I\delta_{k1}$, mass cutoff $M\gg 1$
&\footnotesize
$|\theta|=|\nu-\mu|>1$ & \cite{Ball:2012,Connaughton:2018} \\
\hline
\footnotesize
$K_{ij}=i^{-a}j^{a}+i^{a}j^{-a}$, $F_{ij}=\lambda K_{ij}$, $I_k=0$ & \footnotesize
Brownian kernel $a=\nu=-\mu$, $a\simeq 1^{-}$, $\lambda\ll 1$ & \cite{Brilliantov:2018}
\\
\hline
\footnotesize
$K_{ij}=i^{\mu}j^{\nu}+i^{\nu}j^{\mu}$, $F_{ij}=\lambda K_{ij}$, $I_k=0$ & \footnotesize
$1<\theta=\nu-\mu\simeq 2^{-}$ ($1/2<a\simeq 1^{-}$) &
\\
&\footnotesize and $\lambda<\lambda_c(\theta)\ll 1$ & \cite{Brilliantov:2018}
\\
\hline
\end{tabular}}
\end{table}
%
The manuscript, which tries to detail the dynamics of this model, is organized as follow. In section \ref{sec_model_gen}, we describe the general model and write the set of Smoluchowski master equations for the dynamics of each cluster
density. These equations can be reduced to a set of self-consistent integral equations, depending
solely on the solution for the monomer density which satisfies a non-linear integral equation of Volterra type. The structure of these integrals is dependent on the eigenstates of the main coefficient matrix with eigenvalues in the complex plane. As a first simple example, we demonstrate in section \ref{sec_model_sawtooth} that the monomer density can only be positive and develops a complex time signal. In section \ref{sec_harmonic}, we consider the approximation case of a single harmonic oscillator with negative damping for representing the eigenstate whose eigenvalue has the largest positive real part and we rewrite the Volterra equation for the monomer density as a second order non-linear ordinary differential equation (ODE). The monomer density reaches either a constant value in the long time limit or displays steady oscillations, depending on the input parameters of the model. The frequency of these oscillations is independent on the current and can be computed perturbatively around the oscillations onset. In section \ref{sec_model_N},
we extend this method to the main model in order to obtain a more precise value of the oscillation frequency, and in the conclusion we explain why this oscillating phenomenon is related to the Liénard self-oscillations observed in non-linear mechanical systems with a negative damping.

\section{Main model}\label{sec_model_gen}
%
\begin{figure}[hb]
\centering
\includegraphics[scale=0.8,angle=0,clip]{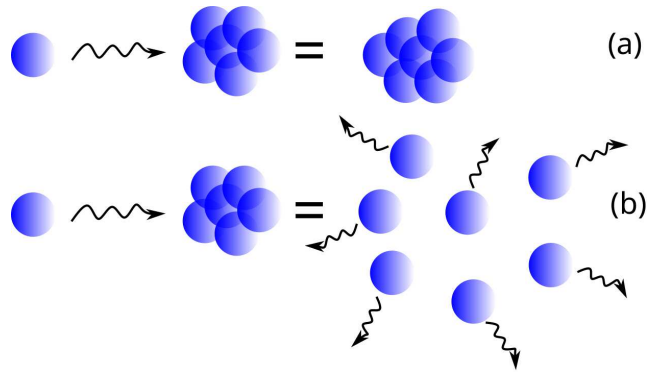}
\caption{\label{fig1_model}
Aggregation and fragmentation process for the main model. A one-particle collides and aggregates with a $k$-particle cluster of size $k$ with probability $\ak_k$, or fragment it into $k$ individual particles with probability $\lambda\ak_k$.}
\end{figure}
%
We consider a set of clusters containing $k$ particles, $1\le k\le N$, each with a density $n_k$. Only one-particle clusters react with other clusters of size $k\ge 1$, see figure \ref{fig1_model}. A particle can aggregate with a cluster of size $k$ to form a cluster of size $k+1$ with probability $\ak_k$.
Or it can shatter this cluster entirely into $k$ single particles with probability $\lambda\ak_k$, as considered in previous works \cite{Brilliantov:2018}. Usually it is assumed that $\lambda\ll 1$.
The rates $\ak_k$ grow according to the cluster size $k$, as a power law with exponent
$a$, such that $\ak_k\propto k^a$. A source of monomer particles is finally introduced with rate
$\mu$, and a sink term with rate $-\mu/N$ for the $N$-particle clusters in order to conserve the 
total mass. The Smoluchowski equations therefore read 
\bb\nn
\frac{\partial \nk_1}{\partial t}=
-2\nk_1^2-\sum_{k=2}^{N-1}(1-k\lambda)\ak_k\nk_1\nk_k+N\lambda\ak_N\nk_1\nk_N
+\mu,
\\ \nn
\frac{\partial \nk_k}{\partial t}=
\ak_{k-1}\nk_1\nk_{k-1}-(1+\lambda)\ak_k\nk_1\nk_k,\;2\le k\le N-1
\\ \label{eq_gen_nk}
\frac{\partial \nk_N}{\partial t}=
\ak_{N-1}\nk_1\nk_{N-1}-\lambda\ak_N\nk_1\nk_N-\mu/N.
\ee
This set of equations satisfies the balance relation for the mass current $\partial_t\sum_{k\ge 1}k\nk_k=0$. It is convenient to redefine the cluster densities by introducing the rescaled variables $\vk_k=\ak_k\nk_k$ and a new definition for the time $\tau=\tau(t)$ such as $d\tau=\nk_1dt$. We obtain in this case a series of quasi-linear ODEs
\bb\nn
&\frac{\partial \vk_1}{\partial\tau}=-2\vk_1-\sum_{k=2}^{N-1}(1-k\lambda)\vk_k
+N\lambda\vk_N
+\mu/\vk_1,
\\ \nn
&\ak_k^{-1}\frac{\partial \vk_k}{\partial \tau}=
\vk_{k-1}-(1+\lambda)\vk_k,\;2\le k\le N-1
\\ \label{eq_gen_vk}
&\ak_N^{-1}\frac{\partial \vk_N}{\partial \tau}=
\vk_{N-1}-\lambda\vk_N-\mu(N\vk_1)^{-1}.
\ee
%
\subsection{Algebraic analysis}
%
In order to solve formally the system \eref{eq_gen_vk}, we rewrite it in the form $\partial_{\tau}\bvk=A\bvk+\bic$, with $\bvk=(\vk_1,\cdots,\vk_N)$, $A$ the $N\times N$ coefficient matrix of the system, and $\bic=\mu(1/\vk_1,0,\cdots,0,-\ak_N(N\vk_1)^{-1})$ the current vector. The matrix $A$ is given explicitly by
\bb\label{matrix_A}\fl\fl
A=\left( \begin{array}{cccccc}
-2 & -1+2\lambda & -1+3\lambda & \cdots & -1+(N-1)\lambda & N\lambda \\
\ak_2 & -(1+\lambda)\ak_2 & 0 & 0 & \cdots & 0 \\
0 & \ak_3 & -(1+\lambda)\ak_3 & 0 & \cdots & 0 \\
\vdots & 0 & \ak_4 & -(1+\lambda)\ak_4 & 0 & 0 \\
0 & 0 & \hspace{1.5cm}0\hspace{1cm}\ddots & \hspace{2cm}\ddots & 0 & 0 \\
0 & \cdots & 0 & \ak_{N-1} & -(1+\lambda)\ak_{N-1} & 0 \\
0 & 0 & \cdots & 0 & \ak_N & -\lambda\ak_N 
\end{array} \right).
\ee
The general ODE for $\bvk(\tau)$ can be solved formally using the diagonalization of 
$A$. This leads directly to the integral solution
\bb\label{gen_sol}
\bvk(\tau)=\int_0^{\tau}Pe^{D(\tau-\tau')}P^{-1}\bic(\tau')d\tau'+Pe^{D\tau}P^{-1}\bvk(0),
\ee
where $A=PDP^{-1}$, $D$ is the diagonal matrix containing the complex eigenvalues $D=\diag(\xl_1,\cdots,\xl_N)$, and $P$ the matrix of the eigenvectors. As initial condition, we take a system consisting of only monomer particles, $\bvk(0)=(\vk_1(0),0,\cdots,0)$, with $\vk_1(0)=\nk_1(0)=\mass$ and $\mass$ being the total initial mass that we can take to be unity for simplification.
The eigenvalues $\xl_i$ are roots of the polynomial equation
\bb\label{eq_poly}\fl
(\xl+2)\prod_{k=2}^{N-1}(1+\lambda+\xl\ak_k^{-1})
+\sum_{k=2}^{N-1}(1-k\lambda)\prod_{l=k+1}^{N-1}(1+\lambda+\xl\ak_l^{-1})
=\frac{\lambda N}{\lambda+\xl\ak_N^{-1}},
\ee
%
\begin{figure}[hb]
\centering
\includegraphics[scale=0.6,angle=0,clip]{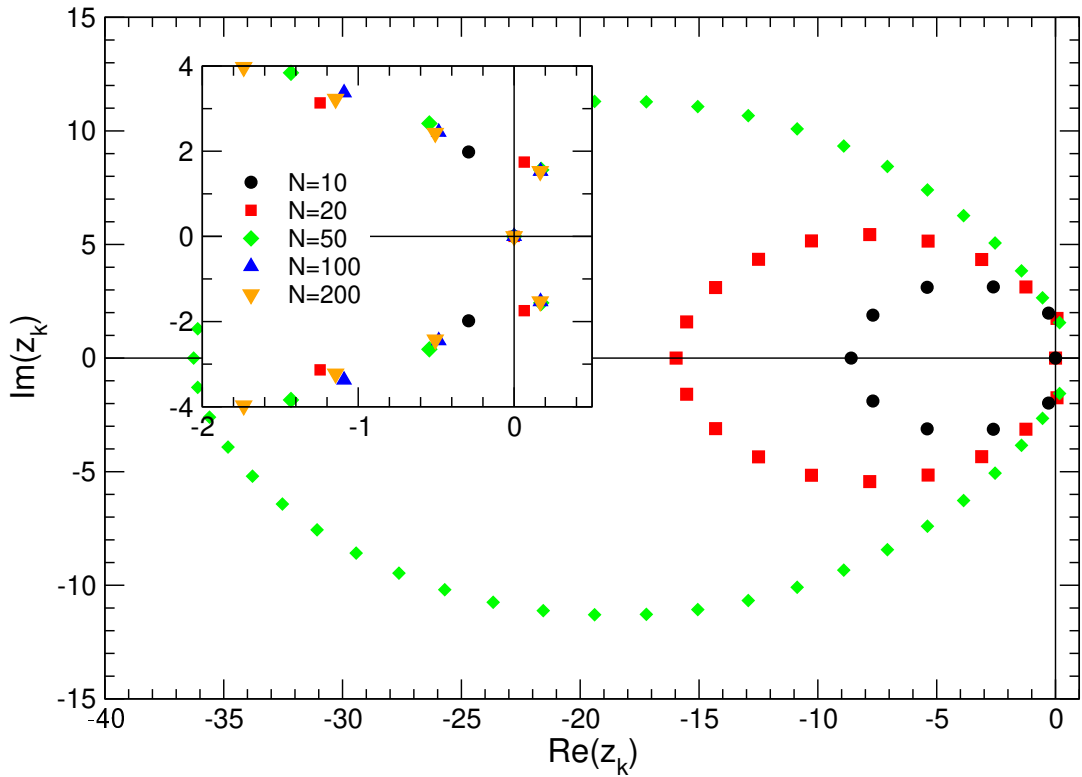}
\caption{\label{fig2_eigen}
Set of eigenvalues of the matrix $A$, with $N=10,20,50,100,200$, exponent $a=0.9$, and
fragmentation factor $\lambda=0.04$. Each eigenvalue comes in conjugate pair in the complex plane, and
the zero eigenvalue is always solution of the polynomial (\ref{eq_poly}). In inset are represented the eigenvalues close to the origin, which converge as $N$ becomes large. Only two of them have a strictly positive real part.}
\end{figure}
%
with the convention that when $k=N-1$ in the sum, the product inside the sum is equal to one.
We can check that the eigenvalue $\xl=0$ is always solution of this polynomial, and corresponds physically to a steady state. In the limit $N\gg 1$, with $\ak_k=k^{a}$, $\lambda>0$ and $a>0$, the product $\prod_{k=2}^{N-1}(1+\lambda+\xl\ak_k^{-1})$ is generally diverging and the secular equation (\ref{eq_poly}) tends to the implicit series
\bb
1=\sum_{k=1}^{\infty}(\lambda k-1)\prod_{l=1}^k
\left (1+\lambda+\xl K_l^{-1}\right )^{-1}.
\ee
This equation has real solutions near the poles $\xl=-(1+\lambda)\ak_k<0$, as well as the
steady state solution $\xl=0$. However it is difficult to extract complex solutions in this
limit, even numerically.
For the simplest case when $N$ is finite and $a=0$, or $\ak_k=1$, the equation \eref{eq_poly} has $N-1$ solutions located on the unit circle of center $-1-\lambda$, with the value $\xl=-\lambda$ excluded, in addition to the trivial solution $\xl=0$ 
\bb\label{eq_eigen_a0}
\xl_k=-1-\lambda+e^{2i\pi k/N},\;k=1,\cdots,N-1,\;{\textrm{and}}\;\xl_0=0.
\ee
All these solutions except the zero solution have strictly negative real part. To obtain
complex solutions with a strictly positive part, $a$ has to be positive, and $N$
sufficiently large.
In figure \ref{fig2_eigen} are represented the complex eigenvalues for different
sizes $N$. The set of solutions deviates from the circle \eref{eq_eigen_a0}, and one observes that
two complex conjugated eigenvalues converge to a finite value with a positive part in the asymptotic limit, see the inset of figure \ref{fig2_eigen}. In particular, this occurs when $N>10$. When $N=10$, all the non-zero eigenvalues have negative real parts, and in this case we do not observe any oscillatory behavior, as we will see in the following. A necessary condition, but a priori not sufficient, to observe oscillations is to have at least one pair of eigenvalues with a strictly positive real part, which occurs already for $N=20$ in figure \ref{fig2_eigen}. A more precise analysis will be made in section \ref{sec_model_N}.
%
%
Each component of the vector $\bvk$ solution of the system (\ref{gen_sol}), and satisfying the previous initial condition, can be expressed as
\bb\label{eq_diff_vk}
\vk_k(\tau)=\mu\int_0^{\tau}\frac{Q_k(\tau-\tau')}{\vk_1(\tau')}d\tau'+
\tilde Q_k(\tau),
\ee
with the kernel functions $Q_k$ and $\tilde Q_k$ defined by the eigenvectors of the matrix $A$
\bb\nn
Q_{k}(\tau)=\sum_{l=1}^{N}P_{kl}
\left (
{(P^{-1})}_{l1}-\frac{K_N}{N}{(P^{-1})}_{lN}
\right )
e^{\xl_l\tau},
\\ \label{eq_Q1}
\tilde Q_{k}(\tau)=\sum_{l=1}^{N}P_{kl}{(P^{-1})}_{l1}e^{\xl_l\tau},
\ee
and initial conditions $Q_{k}(0)=\delta_{k,1}-(K_N/N)\delta_{k,N}$, $\tilde Q_{k}(0)=\delta_{k,1}$. The behavior of $Q_k(\tau)$ or $\tilde Q_k(\tau)$ is dominated after some transit time by the set of conjugate eigenvalues with the largest real part. If this one has strictly positive real and non zero imaginary parts, $Q_k$ and $\tilde Q_k$  will oscillate
and diverge exponentially. Even if the $Q_k$s and $\tilde Q_k$s are exponentially diverging with time, we will show that the density solutions are finite. 
The above equations are all determined by $\vk_1(\tau)$, which satisfies the following self-consistent functional equation
\bb\label{eq_diff_v1}
\vk_1(\tau)=\mu\int_0^{\tau}\frac{Q_1(\tau-\tau')}{\vk_1(\tau')}d\tau'+
\tilde Q_1(\tau),
\ee
and which is usually called a Volterra integral equation of the second kind in the Hammerstein form \cite{Burton:book}. It is in general difficult to obtain an exact solution for any arbitrary function $Q_1$ as the equation \eref{eq_diff_v1} is explicitly a non-linear form and generally non-local. The Laplace transform can be applied to (\ref{eq_diff_v1}) since it involves a convolution product, however this does not give any more information as there is no useful relations
between the Laplace transforms of the function $\vk_1(\tau)$ and its reciprocal ${\vk_1}^{-1}(\tau)$.
%
%
\begin{figure}[hb]
\centering
\includegraphics[scale=0.5,angle=0,clip]{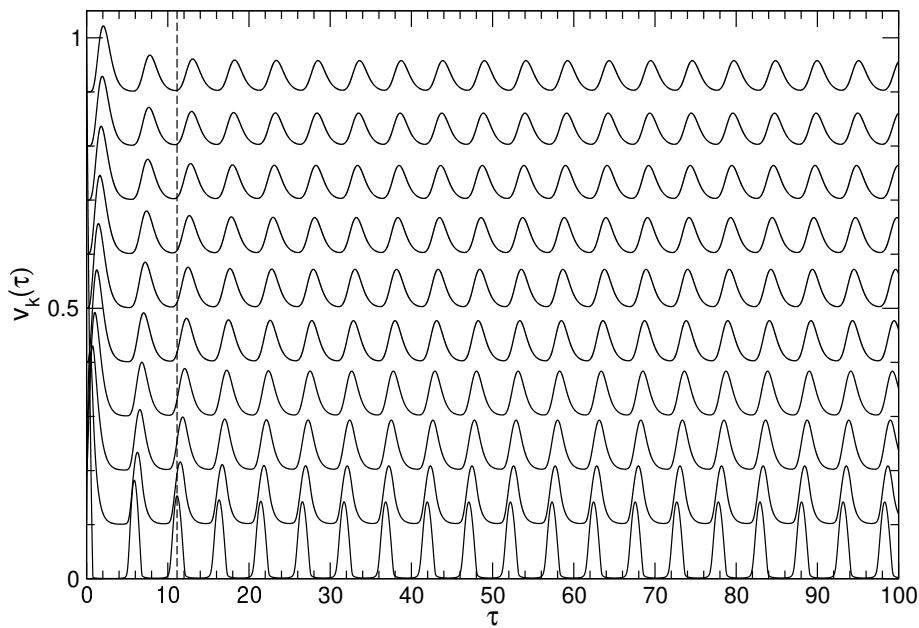}
\caption{\label{fig3_vk}
Numerical solution of the system \eref{eq_gen_vk} for the parameters $N=100$, $a=0.9$,
$\lambda=0.04$, and $\mu=2.10^{-3}$. The successive oscillating densities from $\vk_{1}(\tau)$ (bottom)
to $\vk_{10}(\tau)$ (top) are shown
and shifted upwards for more clarity. The vertical dashed line shows the progressive phase shift between the densities.
}
\end{figure}
%
\subsection{Steady state solution}\label{subsec_constant}
Before analyzing the time evolution of the equation $\partial_{\tau}\bvk=A\bvk+\bic$, we can
search for constant solutions in the long time limit. Since $A$ is singular, solving directly
\eref{eq_gen_vk} with $\partial_{\tau}\vk_i=0$ does not lead to any definite answer. 
We consider instead the solution of the equation $A\bvk=0$, which is a one-dimensional vector, and after some computation we find that the kernel $\ker(A)$ is generated by the vector
\bb
\bvk=\vk_1\left (
1,(1+\lambda)^{-1},(1+\lambda)^{-2},\cdots,(1+\lambda)^{2-N},\lambda^{-1}(1+\lambda)^{2-N}
\right )^{\top},
\ee
where $\vk_1$ is any arbitrary constant. It is easy to show that if we add in the last component of this vector the term $-\mu(N\lambda\vk_1)^{-1}$, then the new vector satisfies $A\bvk+\bic=0$. In order to determine
$\vk_1$, we use the mass conservation equation $\sum_{k=1}^Nk\vk_k/\ak_k=1$ and obtain a second 
order equation for $\vk_1$
\bb\label{eq_v1_N}
\sum_{k=1}^{N-1}\frac{k\vk_1}{\ak_k(1+\lambda)^{k-1}}+\frac{N\vk_1}{\ak_N\lambda(1+\lambda)^{N-2}}
-\frac{\mu}{\ak_N\lambda\vk_1}=1.
\ee
This equation always possesses only one positive solution $\vl$, and in particular, for $N\gg 1$, we obtain that $\vl^{-1}\simeq\sum_{k\ge 1}k^{1-a}/(1+\lambda)^{k-1}$, with the notation $\ak_k=k^a$. This
solution is independent of $\mu$ in this limit. The densities $n_k$ then satisfy the equilibrium values
\bb
n_k=\frac{k^{-a}}{(1+\lambda)^{k-1}}\vl\simeq \frac{e^{-\lambda (k-1)}}{k^a}\vl.
\ee
%
%
\subsection{Numerical analysis}
%
Numerically, we can solve the system of equations \eref{eq_gen_vk} with a 4$^{th}$ order Runge-Kutta algorithm for $N$ different densities
$\vk_k$ as a function of the effective time $\tau$. In the following we consider only the time $\tau$
since the real time $t$ is an increasing and monotonic function of $\tau$. Indeed $n_1$ is found to be always strictly positive. In figure \ref{fig3_vk} is represented the first 10 densities $\vk_{k=1,\cdots,10}$ when $N=100$ for the set of parameters $a=0.9$, $\lambda=0.04$, and $\mu=0.002$. They all display collective oscillations, with a retarded phase shift between each consecutive curve. For this
number $N$, the largest pair of eigenvalues of the matrix $A$ is found to be approximately equal to
$z_{max}^{\pm}\simeq 0.169\pm 1.531 i$, and has therefore a strictly positive real part.
All the solutions of system \eref{eq_gen_vk} are oscillating at a frequency $\Omega$ estimated after
Fourier transform of the $\vk_1$ signal to be equal to $\Omega\simeq 1.226$, which is lower than the
imaginary part $\Im(z_{max}^+)$. We therefore can not relate directly the observed frequency to $\Im (z_{max}^{+})$, which is a natural frequency of the problem, and a more detailed analysis has to be performed in the next sections.

%
\section{Simplified model}\label{sec_model_sawtooth}
%
In order to investigate the previous ODE (\ref{eq_diff_v1}), and show in particular
that $\vk_1$ is always a positive quantity with time,
we consider a very simplified kernel version of $Q_1(\tau)$ and $\tilde Q_1(\tau)$ which is represented by the general sawtooth or triangular function with equidistant time steps $\tau_{n\ge 0}$ such that $T=\tau_{n+1}-\tau_n$ with $\tau_0=0$, and increasing slope coefficients $\alpha_n>0$
\bb\nn
Q_1(\tau)=1-\alpha_0\tau,\;0\le\tau\le T
\\ \label{mod_Q}
Q_1(\tau)=Q_1(\tau_n)+(-1)^{n+1}\alpha_n(\tau-\tau_n),\;\tau_n\le\tau\le\tau_{n+1},
\ee
where the extrema are given by $Q_1(\tau_n)=1+T\sum_{k=0}^{n-1}(-1)^{k+1}\alpha_k$.
Replacing $Q_1$ and $\tilde Q_1$ in equation (\ref{eq_diff_v1}) by (\ref{mod_Q}),
and differentiating with respect to $\tau$, we obtain the following first order and non-local integro-differential equations for each time interval
\bb\nn
\vk_1'(\tau)&=\frac{\mu}{\vk_1(\tau)}-\mu\alpha_0\int_{\tau-\tau_1}^{\tau}\frac{d\tau'}{\vk_1(\tau')}
+\mu\alpha_1\int_{\tau-\tau_2}^{\tau-\tau_1}\frac{d\tau'}{\vk_1(\tau')}
-\cdots
\\ \label{eq_ODE_first}
&+(-1)^{n+1}\mu\alpha_n\int_0^{\tau-\tau_n}\frac{d\tau'}{\vk_1(\tau')}+(-1)^{n+1}\alpha_{n}
,\;\tau_n\le\tau\le\tau_{n+1}.
\ee
Deriving a second time with respect to $\tau$, we obtain a set of piecewise second order ODEs
\bb\nn
\vk_1''(\tau)&=-\frac{\mu\vk_1'(\tau)}{\vk_1^2(\tau)}-\frac{\mu\alpha_0}{\vk_1(\tau)}
+\frac{\mu(\alpha_0+\alpha_1)}{\vk_1(\tau-\tau_1)}-\frac{\mu(\alpha_1+\alpha_2)}{\vk_1(\tau-\tau_2)}
+\cdots
\\ \label{eq_ODE}
&+(-1)^{n+1}\frac{\mu(\alpha_{n-1}+\alpha_n)}{\vk_1(\tau-\tau_n)},\;\tau_n\le\tau\le\tau_{n+1}.
\ee
Each solution on the intervals $\tau_n\le\tau\le\tau_{n+1}$ depends on the previous solutions $\vk_1(\tau-\tau_k)$ for $k=0,\cdots,n$ and can be recursively computed from the first solution in the interval $0\le\tau\le T$ which satisfies
\bb\label{eq_ODE_0}
\vk_1^2(\tau)\vk_1''(\tau)
=-\mu\left (\vk_1'(\tau)+\alpha_0\vk_1(\tau)\right ).
\ee
Initial conditions are $\vk_1(0)=1$ and $\vk_1'(0)=\mu-\alpha_0$
\footnote{We could also rescale the time variable $s=\alpha_0\tau$ with $\tvk_1(s)=\vk_1(\tau)$ to express \eref{eq_ODE_0} with only one parameter $\kappa=\mu/\alpha_0$, such that in the first time interval $\tvk_1^2(s)\tvk_1''(s)=-\kappa(\tvk_1'(s)+\tvk_1(s))$.}. Solving this system of ODEs
requires the continuity of $\vk_1$ at the interval boundaries: $\vk_1(\tau_1^+)=\vk_1(\tau_1^-)$. There also exists a relation between the first derivatives at the boundaries $\vk_1'(\tau_1^+)=\vk_1'(\tau_1^-)+\alpha_0+\alpha_1$, which can be derived directly from equation \eref{eq_ODE_first}. We use the following generalized conditions at $\tau=\tau_n$:
\bb
\vk_1(\tau_n^+)=\vk_1(\tau_n^-),\;\vk_1'(\tau_n^+)=\vk_1'(\tau_n^-)+(-1)^{n+1}(\alpha_{n-1}
+\alpha_n).
\ee
%
\subsection{Solution in the time interval $0\le\tau\le T$}
In the first interval $0\le\tau\le T$, we assume that the parameter $\kappa=\mu/\alpha_0$ is small. The initial conditions are given by $\vk_1(0)=1$ and $\vk_1'(0)=\mu-\alpha_0<0$, as seen previously. We expect in this case that $\vk_1(\tau)$ decreases initially with $Q_1(\tau)$. The overall exact solution is given by an implicit relation between $\tau$ and $\vk_1$. We first assume that $\vk_1$ can be expressed implicitly through the functional integral
\bb\label{eq_Z}
\tau=\int_{\vk_1(\tau)}^1\frac{v}{Z(v)}dv,
\ee
where $Z(v)$ is an unknown function to be determined. By differentiating the previous expression with respect to $\tau$, we have the identity $Z(\vk_1(\tau))=-\vk_1(\tau)\vk_1'(\tau)>0$. Expressing $\vk_1'(\tau)$ and $\vk_1''(\tau)$ in the equation \eref{eq_ODE_0} as functions of $Z(\vk_1(\tau))$ and $\vk_1(\tau)$, we obtain a new ODE of reduced order
\bb
Z'(v)-\frac{Z(v)}{v}=\mu\left (\frac{1}{v}-\frac{\alpha_0 v}{Z(v)}\right ),
\ee
where $Z'(v)$ is the derivative of $Z$ with respect to $v$ and not $\tau$. The initial condition at $\tau=0$ is replaced by $Z(1)=\alpha_0-\mu>0$, according to the previous values of $\vk_1(0)$ and $\vk_1'(0)$. We then perform the substitution $W(v)=(Z(v)+\mu)/v$ in order to simplify even further the previous ODE
\bb\label{eq_W}
W'(v)=\frac{\mu\alpha_0}{\mu-vW(v)}.
\ee
This equation belongs to the class of Abel ODEs of the second kind \cite{book:Polyanin2003,book:Murphy}. A way to resolve it is to use an inverse transformation, solving $v$ as a function of $W$ instead \cite{Cheb-Terrab:2003}. This yields the linear equation
\bb
\mu\alpha_0 v'(W)=\mu-Wv(W),
\ee
whose solutions $v(W)$ are simply given by the integrating factor 
\bb\label{eq_vWsol}
v\exp\left (\frac{W^2}{2\mu\alpha_0}\right )-\alpha_0^{-1}\int_0^{W}
\exp\left (\frac{u^2}{2\mu\alpha_0}\right )du=C,
\ee
which determines $W$ and $Z$ as a function of $v$ implicitly. $C$ is a constant
which can be evaluated by considering the identity $W(\vk_1(0))=\alpha_0$ at $\tau=0$. This
gives the following value of $C$ and its expansion for $\kappa=\mu/\alpha_0$ small
\bb
C=e^{1/2\kappa}-\int_0^{1}\exp\left (\frac{u^2}{2\kappa}\right )du
\simeq (1-\kappa)e^{1/2\kappa}.
\ee
A similar expansion for (\ref{eq_vWsol}) can be performed when $\kappa$ is small,
assuming that $\mu$ is small compared to $W$
\bb
\left (v-\frac{\mu\alpha_0}{W(v)}\right )
\exp\left (\frac{W^2(v)}{2\mu\alpha_0}\right )\simeq C.
\ee
We can then obtain a further approximate value for $Z(v)$, assuming that $Z(v)$ always stays positive
\bb
Z(v)\simeq v\sqrt{2\mu\alpha_0\log\left (C/v\right )}-\mu.
\ee
This implies finally that the density $\vk_1(\tau)$ is given implicitly by
\bb\label{eq_sol_v1}
\tau\simeq \int_{\vk_1(\tau)}^1
\frac{vdv}{v\sqrt{-2\mu\alpha_0\log\left (e^{-1/2\kappa}v^{}\right )}-\mu}.
\ee
In equation (\ref{eq_Z}), we have assumed that $\vk_1(\tau)\le 1$ is decreasing and that the
function $Z(v)$ has to be strictly positive, otherwise $\tau$ may not be defined if $Z$ 
vanishes and becomes negative. For small values of $\tau$, we can approximate \eref{eq_sol_v1}
with $Z(v)\simeq \alpha_0 v$ and therefore obtain that $\vk_1(\tau)\simeq 1-\alpha_0\tau$, which follows the initial 
behavior of the function $Q_1$ and decreases as $\tau$ increases.
However, for $\tau$ large, the inspection of 
equation (\ref{eq_sol_v1}) implies that $\vk_1(\tau)$ should reach asymptotically from above a limiting value $v^*>0$ determined by the singularity of $Z$, and solution of $Z(v^*)=0$:
\bb
v^*\sqrt{-2\mu\alpha_0\log\left (e^{-1/2\kappa}{v^*}^{}\right )}=\mu.
\ee
In the limit of $\kappa$ small, we obtain that $v^*\simeq \kappa+\kappa^2\log\kappa$.
As $\tau$ increases, the integrand becomes indeed larger when $v$ approaches $v^*$ from above. This is equivalent to implying that $\vk_1(\tau)\rightarrow v^*>0$ when $\tau\gg 1$, with $T$ or $\alpha_0$ 
large.
\subsection{Numerical solution in the other intervals $\tau_n\le\tau\le\tau_{n+1}$}
%
We study here the global solutions \eref{eq_ODE} for the successive time intervals, 
and for different sawtooth geometries such as the slope of each linear parts, $T=1$ being fixed.
%
\begin{figure}[ht]
\centering
\includegraphics[scale=0.6,angle=0,clip]{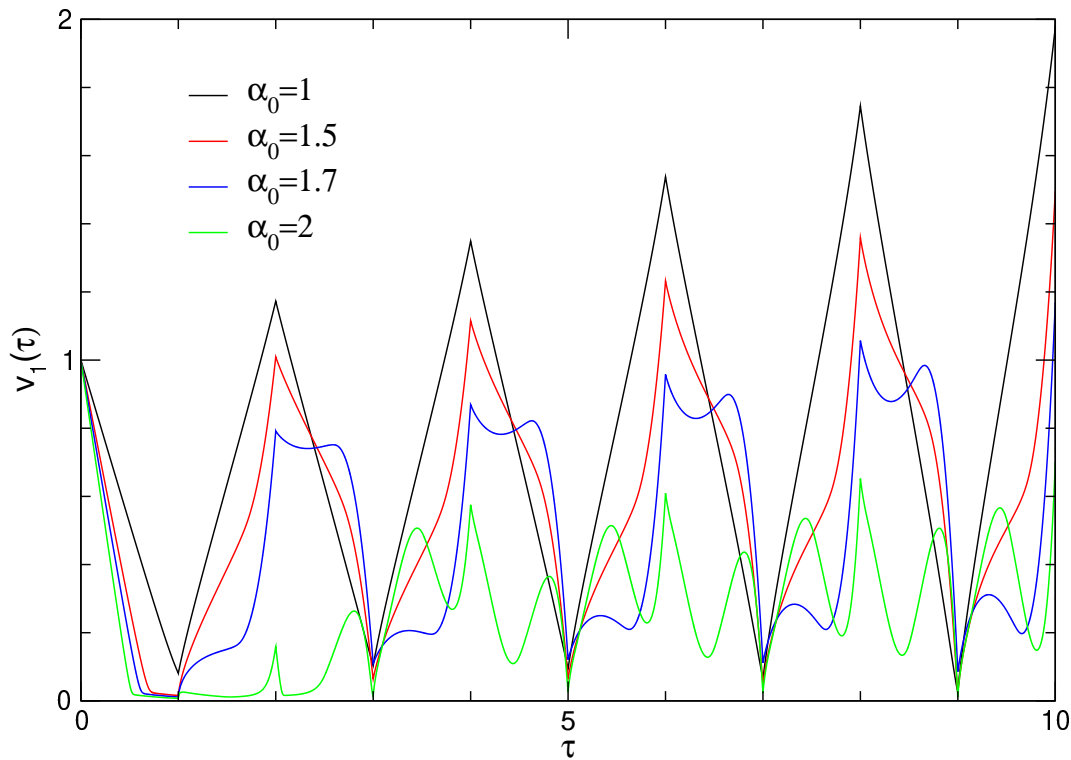}
\caption{\label{fig4_Qtriang}
Solutions of the ODEs \eref{eq_ODE} for a regular sawtooth function $Q_1$
with $\alpha_n=\alpha_0e^{\beta n}$, $\beta=0.1$, $\mu=0.04$, $T=1$, and for different slope 
parameters $\alpha_0$.}
\end{figure}
%
In regard to the time behavior of \eref{eq_Q1}, we consider a diverging amplitude for the oscillating function $Q_1(\tau)$, by taking $\alpha_n=e^{\beta n}\alpha_0$, with $\beta>0$. Figure \ref{fig4_Qtriang}
displays the solutions for different values of $\alpha_0$ with $\beta=0.1$. We have checked that using numerical algorithms for 
solving directly the non-local integral equation \eref{eq_diff_v1}, or solving the ODE \eref{eq_ODE} with the Runge-Kutta-Fehlberg RKF45 method via Maple, leads to the same results.
For small $\alpha_0$, $\vk_1$ follows the sawtooth behavior of $Q_1$ but stays positive.
When $\alpha_0$ increases, $\vk_1$ develops into a more complex periodic signal. It also takes very small values at times $\tau_k=(2k+1)T$, and its maximum amplitude is also smaller as additional frequency modes appear in the signal. It is more difficult to obtain solutions when $\alpha_0>2$, as $\vk_1$ takes very small values in time intervals that cannot be resolved precisely with the algorithms. In summary, $\vk_1$ is periodic and stays positive when we apply the approximate sawtooth kernel function $Q_1$ which is a diverging oscillating function with large negative and positive values. However additional modes appear in the time Fourier spectrum as $\alpha_0$ increases, and which seem to be harmonics of the fundamental frequency $\pi/T$.
%
%
\section{Model with a driven harmonic oscillator and negative damping parameter}
\label{sec_harmonic}
%
%
In this section, we replace the kernels $Q_1(\tau)$ and $\tilde Q_1(\tau)$ by a function representing
an oscillator characterized by a negative damping parameter. The idea is that to approximate
$Q_1(\tau)$ and $\tilde Q_1(\tau)$ defined in equation \eref{eq_Q1} by a simple cosine function of frequency $\omega$ 
multiplied by a diverging exponential factor at rate $\damp$. The quantities $\omega$ and $\damp$ can be viewed as the imaginary and real parts of the largest eigenvalue $z^+_{max}$. Although this approximation is correct asymptotically, it is not the case at small times since all the other eigenstates will give a finite contribution. However we expect to extract enough information about the physics of the system. We also introduce a phase shift $\varphi\in ]-\pi/2,\pi/2[$ such that
\bb\label{eq_Q1_osc}
Q_1(\tau)=e^{\damp\tau}\frac{\cos(\omega\tau-\varphi)}{\cos\varphi},\;\cos\varphi\ne 0,
\ee
with initial condition $Q_1(0)=1$, independent of the phase. The phase diagram is actually symmetrical by the reflection operation $\varphi\rightarrow-\varphi$, so that we will restrict afterwards the analysis to the case when $\varphi\ge 0$.
The derivative of this function at the origin is equal to $Q_1'(0)=\damp+\omega\tan\varphi>0$. Substituting \eref{eq_Q1_osc} into equation \eref{eq_diff_v1}, we obtain
\bb
\vk_1(\tau)=\frac{\mu}{\cos\varphi}\int_0^{\tau}\frac{e^{\damp(\tau-\tau')}
\cos(\omega(\tau-\tau')-\varphi)}{\vk_1(\tau')}d\tau'
+e^{\damp\tau}\frac{\cos(\omega\tau-\varphi)}{\cos\varphi}.
\ee
We can use the fact that $Q_1''-2\gamma Q_1'+(\gamma^2+\omega^2)Q_1=0$ to eliminate the dependence in $Q_1$ by considering the first and second derivative of the previous expression. After combining the different terms we finally obtain a second order ODE independent of the non-local integral
\bb\label{eq_diff_v2}
\vk_1''(\tau)-2\damp \vk_1'(\tau)+(\damp^2+\omega^2)\vk_1(\tau)
=-\mu\frac{\vk_1'(\tau)}{\vk_1^2(\tau)}-(\damp-\omega\tan\varphi)\frac{\mu}{\vk_1(\tau)}.
\ee
The initial conditions are $\vk_1(0)=1$ and $\vk_1'(0)=\damp+\omega\tan\varphi+\mu$. Notice that we
can remove the dependence in $\mu$ by substituting $\vk_1\rightarrow\sqrt{\mu}\vk_1$.
The left hand side is the linear ODE of an harmonic oscillator that $Q_1(\tau)$ satisfies, with frequency
$\omega$ and negative damping parameter $-\damp$, and the right hand side are non-linear terms.
This equation has a time-independent solution in the long time limit provided that $\tan\varphi>\damp/\omega$ and which is equal to
\bb\label{eq_v1_st}
\vl=\left (\mu\frac{\omega\tan\varphi-\damp}{\damp^2+\omega^2}
\right )^{1/2}.
\ee
In the opposite case, $\tan\varphi\le\damp/\omega$, the solution $\vk_1(\tau)$ 
decays to zero exponentially with $\tau$, since for small $\vk_1$ the two terms on the right hand 
side of equation \eref{eq_diff_v2} are dominant, and we have instead
\bb\label{eq_diff_small}
\vk_1'(\tau)\simeq -(\damp-\omega\tan\varphi)\vk_1(\tau).
\ee
This yields $\vk_1(\tau)\sim e^{-(\damp-\omega\tan\varphi)\tau}\rightarrow 0$. To study the stability of the solution \eref{eq_v1_st}, we introduce a correction $\veps(\tau)$ to the steady state solution $\vk_1(\tau)=\vl+\veps(\tau)$, and linearize the equation \eref{eq_diff_v2}
\bb\label{eq_diff_w1}
\veps''(\tau)-\left (2\damp-\frac{\mu}{\vl^2}\right )\veps'(\tau)
+2(\damp^2+\omega^2)\veps(\tau)=0.
\ee
The solution $\veps$ will grow if the damping factor is positive, or $2\damp\ge \mu/\vl^2$.
This will occur if the phase $\varphi$ is larger than a critical value $\varphi_c$ which satisfies
\bb
\tan\varphi\ge \frac{3\damp^2+\omega^2}{2\damp\omega}=\tan\varphi_c> \frac{\damp}{\omega}.
\ee
In this case, we observe steady oscillations with a frequency $\Omega$ determined approximately
at this order by
\bb\label{eq_Omega}
\Omega^2
\simeq 2\damp^2+2\omega^2,
\ee
and which is independent of the current $\mu$ as expected. In table \ref{tab:table3} we have summarized the
phase diagram as function of $\varphi$, according to the properties of \eref{eq_diff_w1}.
%
\begin{figure*}[!ht]
\subfloat[]{
\includegraphics[angle=0,scale=0.32,clip]{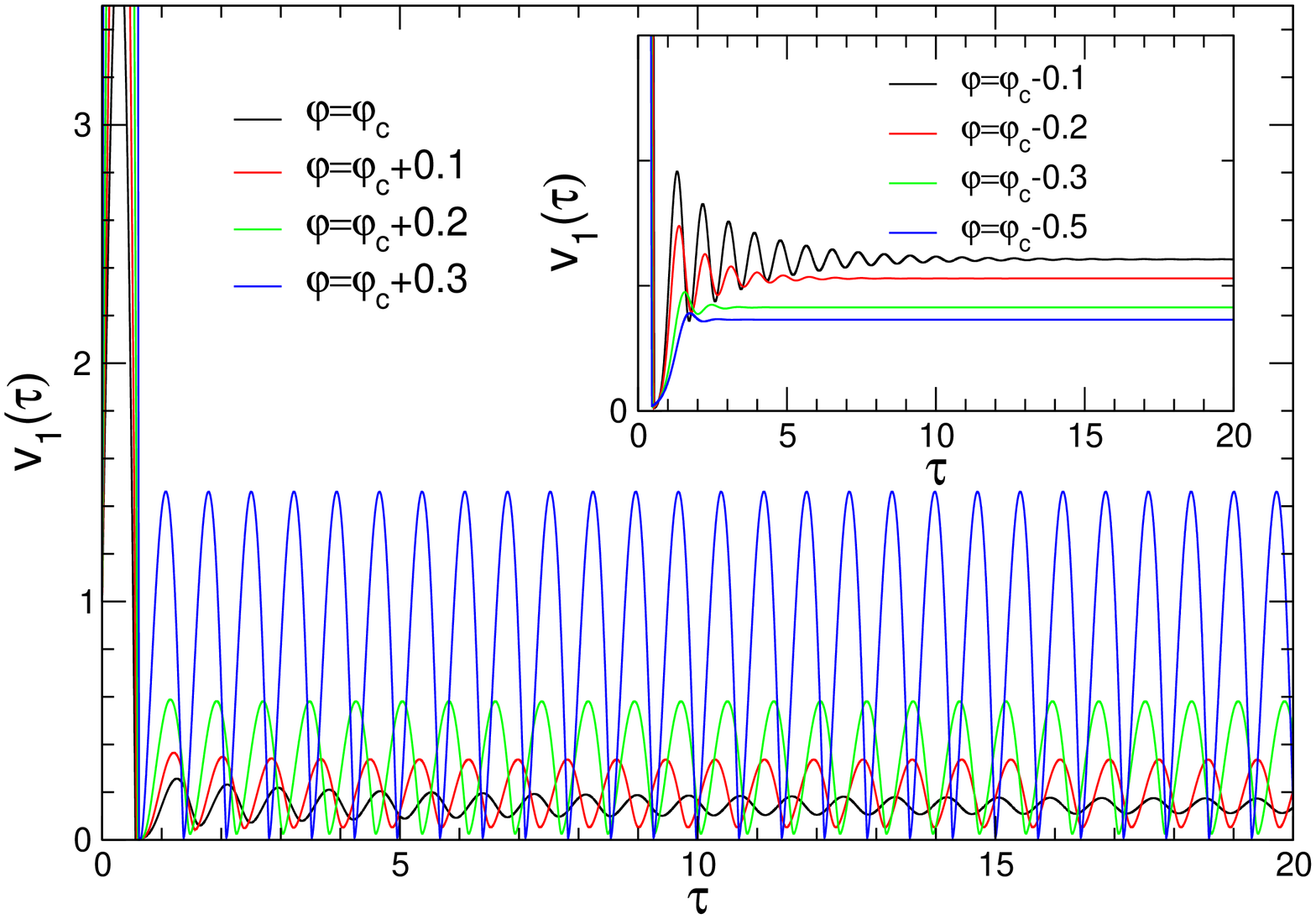}
}
\subfloat[]{
\includegraphics[angle=0,scale=0.32,clip]{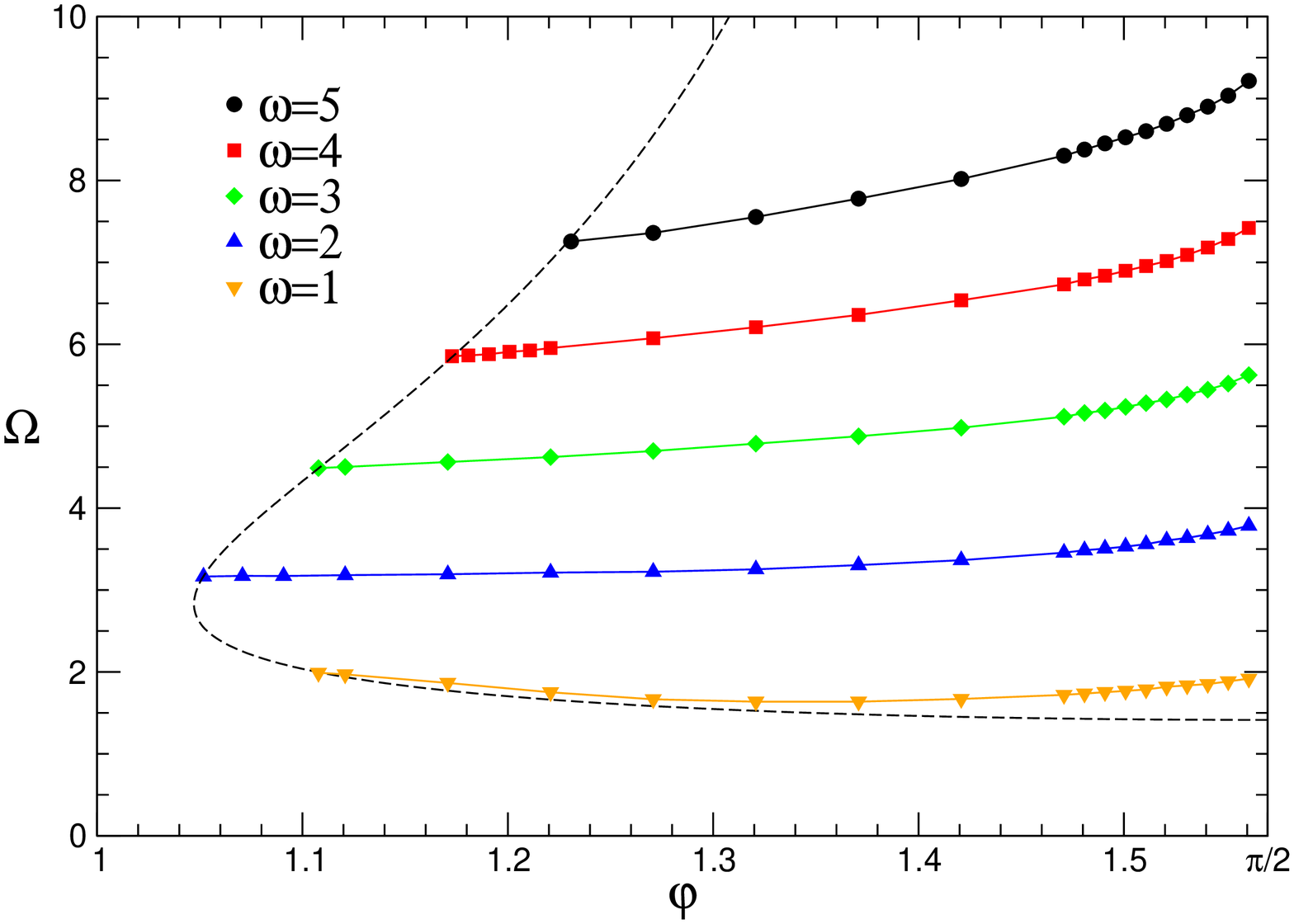}
}
\caption{\label{fig5_Qcos}
(a) Solution of the ODE \eref{eq_diff_v2} for different angles $\varphi$.
The parameters are $\omega=5$, $\damp=1$, and $\mu=0.04$. The critical angle is $\varphi_c=
\arctan(14/5)\simeq 1.228>\arctan(\damp/\omega)=\arctan(1/5)$. 
(b) Oscillation frequency $\Omega$ as function of $\varphi$
for several values of $\omega$, with parameters $\damp=1$, and $\mu=0.04$. 
The dashed line is the threshold $\varphi=\varphi_c$ for which oscillations appear and which delimitates
the domain of existence of $\Omega$. This corresponds to $\Omega^2=2\damp^2+2\omega^2$. 
In the limit $\varphi=\varphi_c=\pi/2$, $\omega\rightarrow 0$ and $\Omega=\sqrt{2}\damp$.}
\end{figure*}
%
\begin{figure*}[!ht]
\subfloat[]{
\includegraphics[angle=0,scale=0.32,clip]{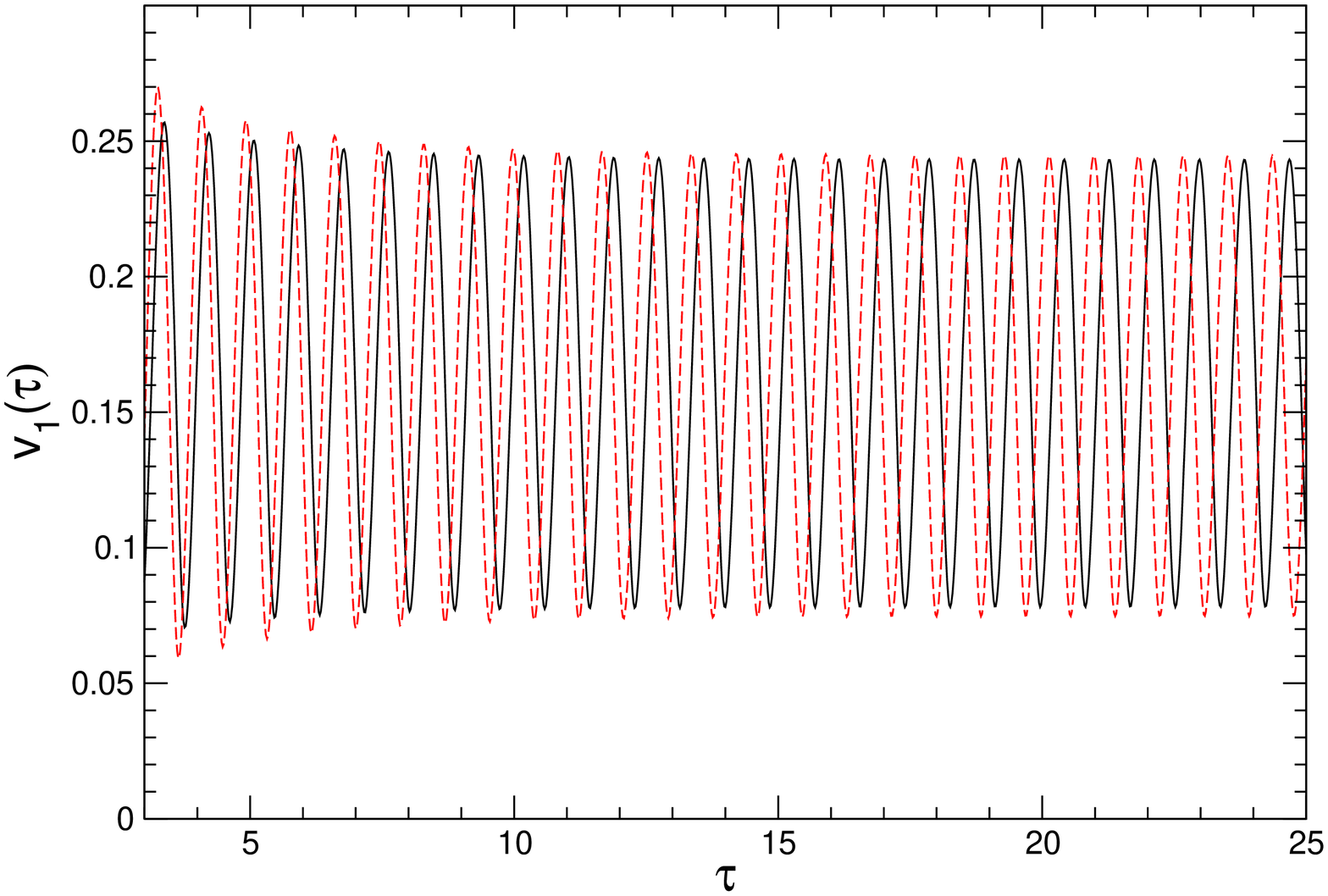}
}
\subfloat[]{
\includegraphics[angle=0,scale=0.32,clip]{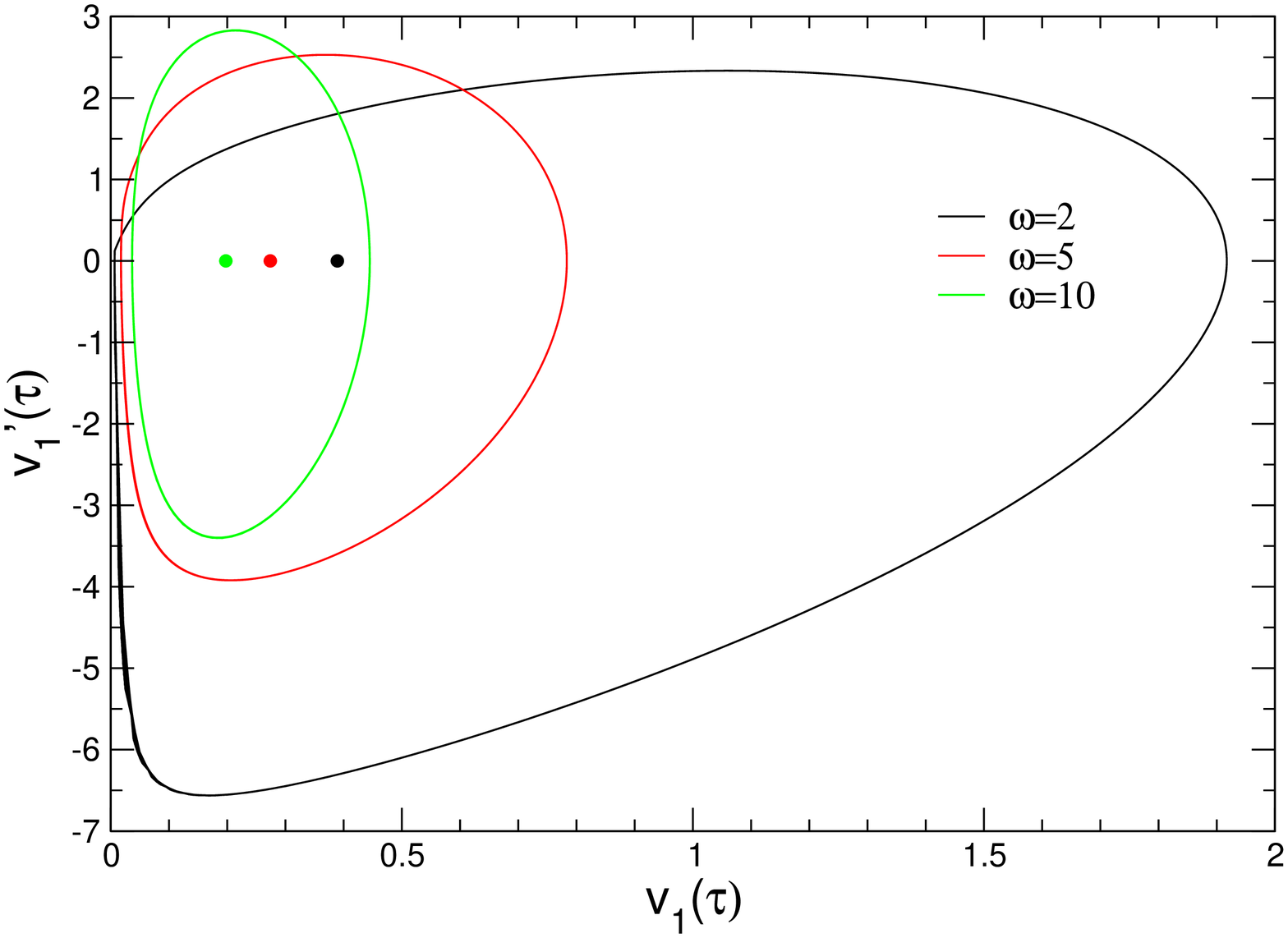}
}
\caption{\label{fig6_cycle}
(a) Approximation corresponding to \eref{eq_diff_w3} (red dashed line) and comparison with the exact
solution \eref{eq_diff_v2} (black line) for $\omega=5$, $\damp=1$, and $\mu=0.04$. $\varphi=\pi/2-0.3$ is chosen to be close to $\varphi_c$: $\delta\varphi=\varphi-\varphi_c\simeq 0.043$.
(b) Oscillation cycles for three different parameters $\omega$, with $\damp=1$, $\varphi=\pi/2-0.1$, and $\mu=0.04$. The dot points correspond to the equilibrium solution $\vl$.}
\end{figure*}
%
%
\begin{table}[h]
\centering
\caption{}
\label{tab:table3}
{\renewcommand{\arraystretch}{1.2}
\begin{tabular}{c|c}
$0\le\tan\varphi\le\damp/\omega$ & $\vk_1(\tau)$ goes to zero exponentially \eref{eq_diff_small}
\\
\hline
$\damp/\omega<\tan\varphi\le(3\damp^2+\omega^2)/2\damp\omega$
& $\vk_1(\tau)$ goes to the equilibrium value $\vl$ \eref{eq_v1_st} 
\\
\hline
$(3\damp^2+\omega^2)/2\damp\omega<\tan\varphi$ &
$\vk_1(\tau)$ oscillates with a finite amplitude and frequency $\Omega$ \eref{eq_Omega}
\end{tabular}}
\end{table}

In figure \ref{fig5_Qcos}(a) is represented numerically the time evolution of $\vk_1$ as function of different values of the phase $\varphi$. There are two domains according to whether $\varphi$ is lower or greater than $\varphi_c$, for which $\vk_1$ reaches asymptotically the equilibrium value
$\vl$ or self-oscillates respectively. The oscillation frequency $\Omega$ can be extracted by Fourier transform of the time
signals and represented as a curve function of $\varphi$ for different values of $\omega$, see figure \ref{fig5_Qcos}(b). There is no simple relation between $\Omega$ and the other parameters, except
at the critical angle $\varphi_c$ for which $\Omega=\sqrt{2\damp^2+2\omega^2}$, see equation \eref{eq_Omega}. 

Near $\varphi_c$, we can assume that the oscillation amplitude is small compare to
the constant solution $\vl$. The previous linear expansion \eref{eq_diff_w1} shows that the damping factor $2\damp-\mu/\vl^2$ must be positive in order to generate oscillations. However, this expansion leads to an unstable solution as $\veps$ diverges with time. We can study the next terms
of the expansion and verify whether these terms stabilize the oscillation amplitude. If we expand \eref{eq_diff_v2} up to the third order in $\veps$, we obtain a non-linear equation
with a time dependent damping factor
\bb\nn
\veps''(\tau)-\left (2\damp-\frac{\mu}{\vl^2}+2\frac{\mu}{\vl^3}\veps(\tau)-3\frac{\mu}{\vl^4}
\veps^2(\tau)\right )\veps'(\tau)
\\ \label{eq_diff_w3}
+(2\damp^2+2\omega^2)\veps(\tau)+(\damp-\omega\tan\varphi)\frac{\mu}{\vl^3} \veps^2(\tau)-(\damp-\omega\tan\varphi)\frac{\mu}{\vl^4} \veps^3(\tau)=0.
\ee
We have checked that this ODE displays stable oscillations close to the
exact solution for $\varphi\gtrapprox \varphi_c$, see figure \ref{fig6_cycle}(a). 
These oscillations are not symmetric, and their amplitude depends on $\omega$, see figure \ref{fig6_cycle}(b). For example, as $\omega$ decreases, the amplitude increases.
Restricting the expansion to the second order would still lead to 
non stable oscillations, due in particular to the presence of the linear term $\veps(t)$ in the damping
coefficient, in factor of $\veps'(t)$. The stability comes instead from the quadratic term $\veps^2(t)$ in the same damping coefficient and from the last term in $\veps^3(t)$. 
To analyze the stability of \eref{eq_diff_w3}, we consider an expansion in the small
parameter $0<\tan\varphi-\tan\varphi_c\simeq \delta\varphi\ll 1$. The technique we use in the following is described in the reference book \cite{book:Mook}
and can be applied to different ODEs with negative damping containing a linear or quadratic amplitude in addition to the constant term. This is usually useful to study self-excited oscillators.
We first simplify the previous ODE in order to reduce the number of parameters. 
We chose to rescale the amplitude $\veps(\tau)=c\tveps(\tau)$, with $c=\vl\omega\damp\delta\varphi/(\damp^2+\omega^2)$, so that
\bb\label{eq_diff_eps}
\tveps''(\tau)+\Omega_0^2\tveps(\tau)=
4\damp\epsilon\left (
1+\tveps(\tau)-\frac{3\epsilon}{2}\tveps^2(\tau) \right )\tveps'(\tau)+\frac{\epsilon}{2}\Omega_0^2 
\left (\tveps^2(\tau)-\epsilon\tveps^3(\tau) \right ),
\ee
with $\Omega_0^2=2(\damp^2+\omega^2)$, and $\epsilon=\omega\damp\delta\varphi/(\damp^2+\omega^2)\ll 1$ the rescaled small parameter. This yields in particular $c=\epsilon\vl$.
When $\epsilon=0$, we can write the general solution \eref{eq_diff_eps} as 
$\tveps(\tau)=\amp\cos(\Omega_0 \tau+\beta)$. When $\epsilon\ne 0$, the general solution can
still be expressed in the same form, provided the coefficients $\amp$ and $\beta$ are time dependent.
Since we have two functions satisfying one ODE, we can chose a constraint
between $\amp$ and $\beta$. It is usually convenient to impose the identity 
$\tveps'(\tau)=-\amp\Omega_0\sin(\Omega_0 \tau+\beta)$. This implies in particular that
\bb\label{eq_const}
\amp'(\tau)\cos\phi-\amp\beta'(\tau)\sin\phi=0,
\ee
where we used the notation $\phi=\Omega_0\tau+\beta(\tau)$. Equation \eref{eq_diff_eps} can then
be formally rewritten as 
\bb\label{eq_f1}
\tveps''(\tau)+\Omega_0^2\tveps(\tau)=
\epsilon\func (\tveps,\tveps')=
\epsilon\func (\amp\cos\phi,-\amp\Omega_0\sin\phi),
\ee
with function $\func$ defined by
\bb\label{eq_f2}
\func (\tveps,\tveps')=4\damp\left (
1+\tveps-\frac{3\epsilon}{2}\tveps^2 \right )\tveps'+
\frac{1}{2}\Omega_0^2 
\left (\tveps^2-\epsilon\tveps^3 \right ).
\ee
After computing the second derivative $\veps''(\tau)$ from the expression $\veps'(\tau)=-\amp\Omega_0\sin\phi$, we obtain
\bb
\tveps''(\tau)+\Omega_0\tveps(\tau)=-\amp'(\tau)\Omega_0\sin\phi-\amp\Omega_0\beta'(\tau)\cos\phi.
\ee
The right hand side can be identified with \eref{eq_f1}, and using the relation between $\amp$ and $\beta$, see \eref{eq_const}, we obtain the following set of first order ODEs
\bb\nn
\amp'=-\frac{\epsilon}{\Omega_0}\sin\phi\func(\amp\cos\phi,-\amp\Omega_0\sin\phi),
\\
\beta'=-\frac{\epsilon}{\Omega_0u}\cos\phi\func(\amp\cos\phi,-\amp\Omega_0\sin\phi).
\ee
These equations are identical to \eref{eq_diff_eps}, since no approximation have been made yet. We then assume that $\phi$ varies rapidly compared to the variations of $\amp$ and $\beta$. We can replace in this approximation the previous equations by their time average over a period $2\pi/\Omega_0$, after considering the quantities $\amp'$ and $\beta'$ quasi constant. This yields
\bb\nn
\amp'\simeq-\frac{\epsilon}{2\pi\Omega_0}\int_0^{2\pi}d\phi\sin\phi\func(\amp\cos\phi,-\amp\Omega_0\sin\phi),
\\
\beta'\simeq-\frac{\epsilon}{2\pi\Omega_0u}\int_0^{2\pi}d\phi\cos\phi\func(\amp\cos\phi,-\amp\Omega_0\sin\phi).
\ee
After performing the integrations over $\phi$, we obtain a set of coupled ODEs for the amplitude and 
phase shift
\bb\label{eq_amp}
\amp'=\damp\epsilon(2\amp-\frac{3\epsilon}{4}\amp^3),\;
\beta'=\frac{3}{16}\Omega_0\epsilon^2\amp^2.
\ee
We should notice that the evolution of the amplitude $\amp$ does not depend on the terms in factor of $\Omega_0^2$ in equation \eref{eq_f2}, after averaging over $\phi$, but is determined by the damping terms in factor of $\tveps'$, especially the constant and quadratic term $\tveps^2$. However, the change in frequency depends instead on the cubic term in factor of $\Omega_0^2$. All the terms which are a odd function of $\phi$ do not contribute to \eref{eq_amp}.
The presence of the quadratic term in the damping factor leads to the existence of a
finite equilibrium limit for $\amp$ which converges exponentially to a positive constant $\amp_0$ determined by the non-zero solution of $\amp'=0$ in \eref{eq_amp}, or explicitly $\amp_0=\sqrt{8/(3\epsilon)}$. This leads, after rescaling to the original variable, to the equilibrium amplitude $\veps_0=\sqrt{8\epsilon/3}\vl$, and therefore we can conclude that the effect of the additional quadratic term in the damping factor is to stabilize the oscillations that are generated self-consistently.
The frequency $\Omega$ is modified accordingly after solving \eref{eq_amp} for $\beta'$ in the asymptotic limit, and is shifted by a small amount such that the corrected frequency is given by $\Omega\simeq \Omega_0+\beta'$ or 
$\Omega\simeq\Omega_0(1+\epsilon/2)$. It is, as expected, independent of $\mu$, contrary to the amplitude which is proportional to $\vl\propto\sqrt{\mu}$.
As an example, if we take the parameters from figure \ref{fig5_Qcos}(b), 
$\omega=3$, $\damp=1$ and $\mu=0.04$, we obtain for $\varphi=1.17\gtrapprox\varphi_c\simeq 1.107$ the 
overall estimated amplitude of the signal $2\veps_0\simeq 0.185$, which is close to the numerical value extracted from the analysis of the ODE \eref{eq_diff_eps} or $0.217$, with $\epsilon=0.128$. We also obtain $\Omega\simeq 4.759$, to be compare with $\Omega_0\simeq 4.472$
and the numerical estimation from the Fourier transform of the signal $\Omega\simeq 4.634$.

The present model in this section does not consider all the eigenstates with eigenvalues that have negative real part, since
we have assumed that these states have a negligible contribution as they correspond to the terms in \eref{eq_Q1} that vanish exponentially with time. However their contribution at small times may be relevant to the general solution as the convolution integral equations \eref{eq_diff_vk} and \eref{eq_diff_v1} are non local.
In the following section, we will generalize the previous method to the main model which incorporates
$N$ oscillators.
%
\section{Generalization to a set of harmonic oscillators}\label{sec_model_N}
The generalization of the previous results to the main model of section \ref{sec_model_gen} is important in order to analyze the onset of the self-oscillations, and to extract the value of the frequency observed for example in figure \ref{fig3_vk} and which can not be approximated with a simple oscillator.
We first write a general ODE for $\vk_1$ by removing all dependence in the 
$Q_1$ and $\tilde Q_1$ functions, as well as the convolution integrals. We expect that for $N$ oscillators, $\vk_1$ satisfies a $N^{th}$-order ODE.
As in the previous section, we can write an ODE satisfied by $Q_1$ and $\tilde Q_1$, since they
are a sum of $N$ harmonic oscillators (or $N-1$ precisely because of the presence of the zero eigenvalue eigenstate). From equation \eref{eq_Q1}, it is easy to establish a linear ODE satisfied by $Q_1$ and $\tilde Q_1$ using the characteristic polynomial of the coefficient matrix $A$ whose coefficients are given by
\bb
\det\left (z\mathit{I}_N-A \right) = \sum_{k=0}^Na_kz^k,
\ee
with $a_N=1$ and $a_0=0$. Since the eigenvalues are roots of this polynomial, we have directly
an ODE for $Q_1$ and $\tilde Q_1$ after differentiating \eref{eq_Q1}
\bb
\sum_{k\ge 1}^Na_kQ_1^{(k)}(\tau)=0.
\ee
%
\begin{figure*}[!ht]
\subfloat[]{
\includegraphics[angle=0,scale=0.32,clip]{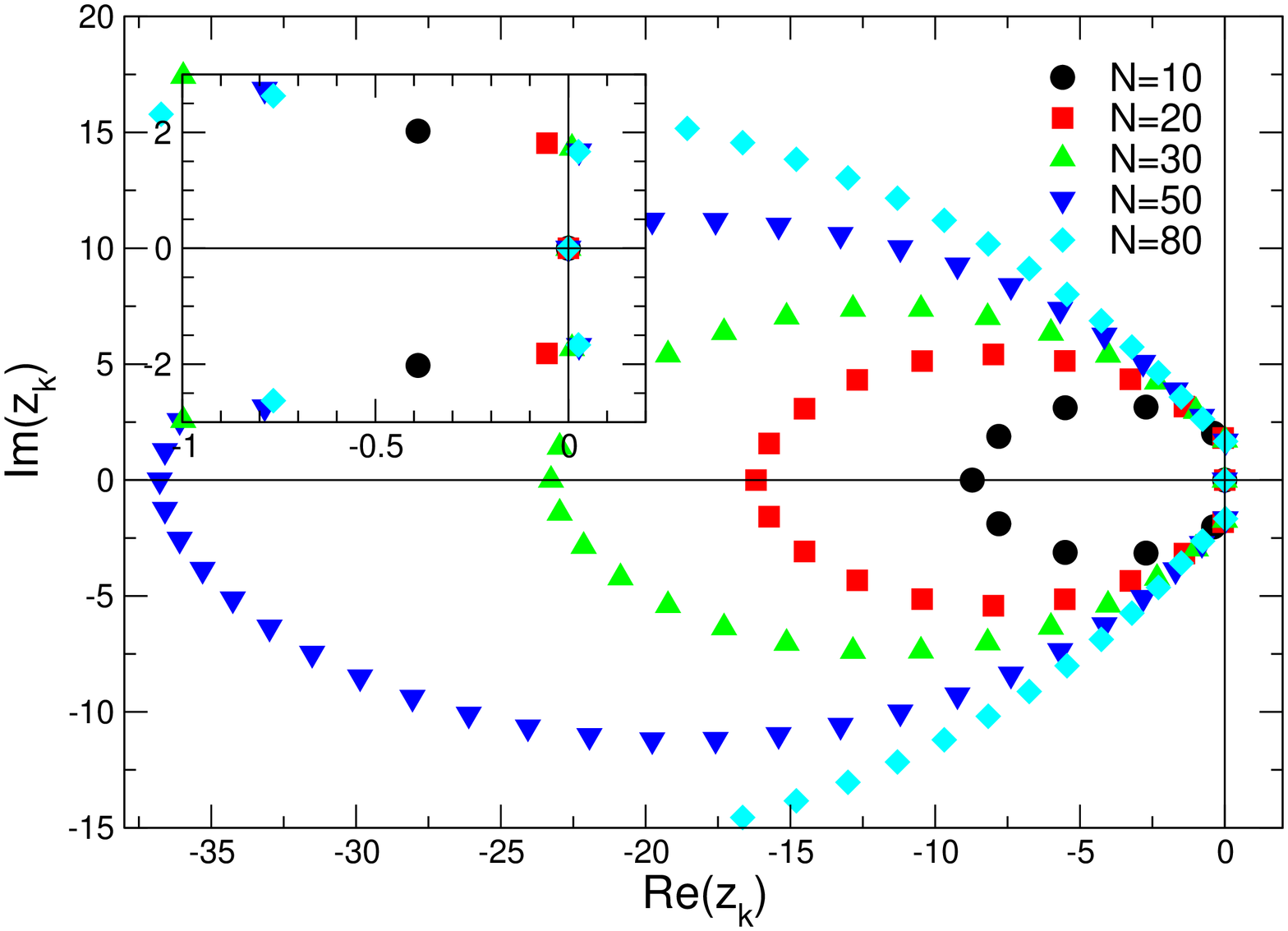}
}
\subfloat[]{
\includegraphics[angle=0,scale=0.32,clip]{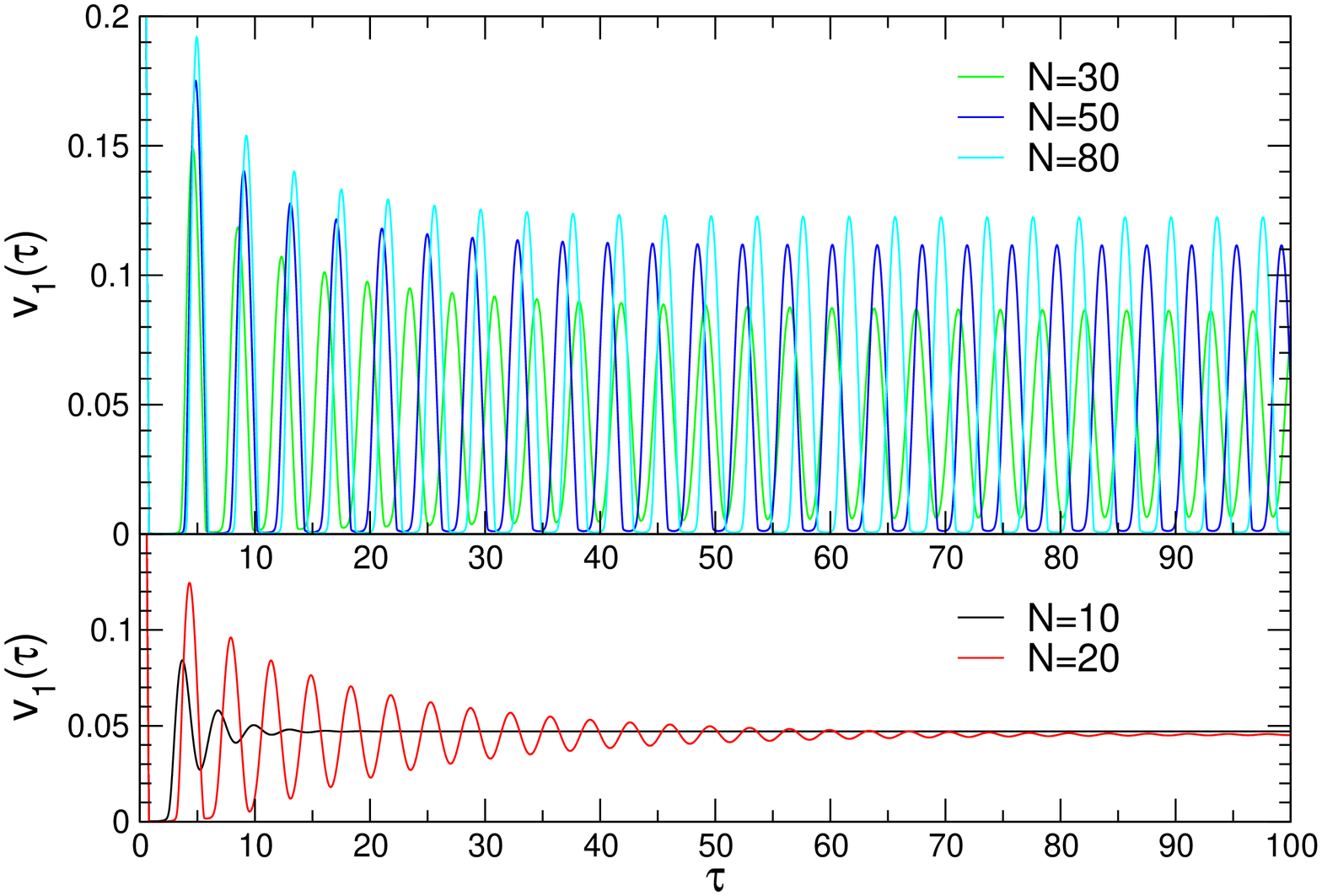}
}
\caption{\label{fig7_eigen_vdiff}
(a) Plot of the complex roots of $R(z)$, equation \eref{eq_R}, for different values of $N$.
The parameters are $\mu=2.10^{-4}$, $\lambda=0.06$ and $a=0.9$. In inset
is represented the roots near the origin. Only the cases $N=30$, $N=50$, and $N=80$ have
a pair of roots with strictly positive real parts.
(b) Time dependence of $\vk_1(\tau)$ for the previous cases. Oscillations are observed for
$N=30,50,80$. For the other cases $N=10,20$, $\vk_1$ reaches the asymptotic limits $\vl\simeq 0.047$
and $\vl\simeq 0.045$ respectively, see equation \eref{eq_v1_N}.
}
\end{figure*}
By differentiating $N$ times the identity \eref{eq_diff_v1}, and multiplying each successive equations by $a_k$, we can eliminate all the time dependence involving $Q_1$ and $\tilde Q_1$, and therefore all the convolution integrals disappear. This yields
\bb\label{eq_diff_v1_gen}
\sum_{k=1}^N a_k\left (
\vk_1^{(k)}-\mu\sum_{l=0}^{k-1}Q_1^{(k-1-l)}(0)\frac{\partial^l}{\partial\tau^l}
\left (\frac{1}{\vk_1}\right )
\right )=0.
\ee
This equation contains only derivatives in $\vk_1$ since $a_0=0$ and also because there
is no contribution in $\vk_1$ coming from the $l=0$ terms in the previous sum as
\bb
\sum_{k=1}^N a_kQ_1^{(k-1)}(0)=0.
\ee
This comes from the fact that one of the factors $\aq_l$
in the sum \eref{eq_Q1}, $Q_1(\tau)=\sum_{l=1}^{N}\aq_le^{z_l\tau}$, vanishes when $l=l_0$, with $l_0$ associated to the zero eigenvalue $z_{l_0}=0$. This has been checked numerically. Otherwise, it is easy to see that the previous equation \eref{eq_diff_v1_gen} would not be consistent with
the existence of the time-independent solution that we found in section \ref{subsec_constant}.
There is therefore no $\vk_1$ term in the ODE, contrary to the case of the single harmonic oscillator that we studied in the previous section. 
Each derivative term of the ODE \eref{eq_diff_v1_gen} can be rearranged by increasing differential order of $\vk_1^{-1}$
\bb\nn\fl
\sum_{k=1}^Na_k\vk_1^{(k)}-\mu\left (\frac{1}{\vk_1}\right )^{'}
\left (a_2Q_1(0)+a_3Q_1'(0)+\cdots+a_NQ_1^{(N-2)}(0)\right )
\\ \nn\fl
-\mu\left (\frac{1}{\vk_1}\right )^{''}
\left (a_3Q_1(0)+a_3Q_1'(0)+\cdots+a_NQ_1^{(N-3)}(0)\right )
-\cdots
\\ \fl
-\mu\left (\frac{1}{\vk_1}\right )^{(N-1)}a_NQ_1(0)=0.
\ee
We can define $(N-1)$ additional coefficients $b_k=\sum_{l=0}^{N-k-1}a_{k+1+l}Q_1^{(l)}(0)$
such that the previous ODE becomes
\bb\label{eq_diff_v1_ab}
\sum_{k=1}^Na_k\vk_1^{(k)}-\mu\sum_{k=1}^{N-1}
b_k\left (\frac{1}{\vk_1}\right )^{(k)}=0.
\ee
We can, as before, eventually 
remove the dependence on $\mu$ by rescaling $\vk_1\rightarrow\sqrt{\mu}\vk_1$.
The complex coefficients $b_k$ can be computed recursively as the inverse powers of the eigenvalues, 
using the factors $\aq_l$. For the first two terms, we have indeed
\bb
b_1=-a_1\sum_{l\ne l_0}\aq_lz_l^{-1},\;
b_2=-a_1\sum_{l\ne l_0}\aq_lz_l^{-2}-a_2\sum_{l\ne l_0}\aq_lz_l^{-1},
\ee
and for the general case $b_k=-\sum_{i=1}^{k}a_i\sum_{l\ne l_0}\aq_lz_l^{-k-1+i}$.
From these results, we can analyze the stability of the steady state solution $\vl$ 
of the equation \eref{eq_v1_N} if we linearize \eref{eq_diff_v1_ab} by setting
$\vk_1=\vl+\wk$, $|\wk|\ll \vl$, which yields
\bb\label{eq_diff_v1_w}
\sum_{k=1}^Na_k\wk^{(k)}+\frac{\mu}{\vl^2}\sum_{k=1}^{N-1}
b_k\wk^{(k)}=0.
\ee
The stability of the solution can be studied by finding the roots of the following polynomial
\bb\label{eq_R}
R(z)=\sum_{k=1}^{N-1}\left (a_k+\frac{\mu}{\vl^2}b_k\right )z^k+a_Nz^N.
\ee
This polynomial is different from the characteristic polynomial of $A$ due to the additional terms $b_k$. The perturbation $w(\tau)$
will grow if at least one complex root of $R(z)$ has a strictly positive real part. In figure  
\ref{fig7_eigen_vdiff}(a) we have represented the roots of $R(z)$ for different $N$, with a fixed set
of parameters. As $N$ increases, one pair of conjugate roots moves into the positive part of the real
axis, and self-oscillations occur, see figure \ref{fig7_eigen_vdiff}(b). The imaginary part of the
root $z_{max}^{+}$ of $R(z)$ with the largest real part gives an estimate of the oscillation frequency.
We find for $N=50$ $z_{max}^+\simeq 0.092+1.454i$, and the Fourier transform of the signal 
of figure \ref{fig7_eigen_vdiff}(b) gives approximately $\Omega\simeq 1.332$ for this value of $N$. $\Im(z_{max}^+)$ is closer to $\Omega$ than the value given by the characteristic polynomial of $A$ which is equal to $\Im(z_{max}^+)\simeq 1.561$.

A more precise analysis of \eref{eq_diff_v1_ab} requires to estimate the contributions from all the non-linear terms. The successive derivatives of $\vk_1^{-1}$ can be ordered as function of the powers
of $\vl^{-1}$. This can be done recursively by using a combinatorial method
\bb\label{eq_diff_v1_ab_diff}\fl
\sum_{k=1}^N
a_k\wk^{(k)}+\mu \sum_{k=1}^{N-1}b_k
\left (\frac{1}{\vl^2}\frac{\wk^{(k)}}{(1+\wk/\vl)^2}-
\frac{1}{\vl^3}\sum_{l_1=1}^{k-1}{k\choose l_1}\frac{\wk^{(l_1)}\wk^{(k-l_1)}}{(1+\wk/\vl)^3}
\right .
\\ \nn\fl
\left .
+\frac{1}{\vl^4}\sum_{l_1=1}^{k-1}\sum_{l_2=1}^{k-l_1-1}{k\choose l_1}{k-l_1\choose l_2}
\frac{\wk^{(l_1)}\wk^{(l_2)}\wk^{(k-l_1-l_2)}}{(1+\wk/\vl)^4}
-\cdots
+\frac{(-1)^{k+1}}{\vl^{k+1}}\frac{k!{\wk^{(1)}}^k}{(1+\wk/\vl)^{k+1}}
\right )
=0.
\ee
This expression is exact. This gives a formal expansion in $\vl^{-1}$.
When $\vl$ is large enough, we can approximate this ODE by retaining only
the first few terms. As in section \ref{sec_harmonic}, we can analyze the stability of the  solution associated to the pair of conjugate roots $z_{max}^{\pm}=\Lambda_0\pm i\Omega_0$ of the polynomial \eref{eq_R} with $\Lambda_0$ assumed in the following to be strictly positive but smaller than $\Omega_0$. We consider the perturbation amplitude $\tveps(\tau)=e^{\Lambda}\cos(\Omega_0\tau+\beta)$, after rescaling $\wk=\sqrt{\mu}\tveps$ and
set $\epsilon=\sqrt{\mu}/\vl$ which is the small parameter of the expansion. 
The quantities $\Lambda=\Lambda_0\tau+\aamp(\tau)$ and $\beta$ depend on time and a perturbative solution can be given as an expansion at the lowest order in $\epsilon$. We also choose a constraint
between the two functions $\Lambda$ and $\beta$ and impose the following identity for the derivative:
$\tveps'(\tau)=e^{\Lambda}(\Lambda_0\cos\phi-\Omega_0\sin\phi)$, where we used the notation 
$\phi=\Omega_0\tau+\beta(\tau)$. This implies the following condition
\bb\label{eq_const_2}
\aamp'(\tau)\cos\phi-\beta'(\tau)\sin\phi=0.
\ee
After some calculations and keeping only the linear terms in $\aamp'$ and $\beta'$, we establish the following successive derivatives of $\tveps$ 
\bb\nn
\tveps^{(k)}(\tau)\simeq e^{\Lambda}\left [
Z_k(\phi)+c_k
\left (\aamp'(\tau)\sin\phi+\beta'(\tau)\cos\phi \right )\right ],
\\ \label{wk}
Z_k(\phi)=\Re\left [({z_{max}^{+}})^ke^{i\phi}\right ],\;
c_k=-\Im \left (
z_{max}^{+}\frac{({z_{max}^{+}})^{k-1}-\Lambda_0^{k-1}}{z_{max}^{+}-\Lambda_0}
\right ).
\ee
We then replace the derivatives of $\tveps$ in the ODE \eref{eq_diff_v1_ab_diff} by their expression from \eref{wk}, and organize the different terms as a series expansion in $\epsilon$. We obtain a single equation for the combination $\aamp'(\tau)\sin\phi+\beta'(\tau)\cos\phi$:
\bb\nn
\left (\aamp'(\tau)\sin\phi+\beta'(\tau)\cos\phi \right )
\sum_{k=1}^{N-1}c_k(a_k+\epsilon^2b_k+{\cal O}(\epsilon^3))\simeq
\\ \nn
\epsilon^3e^{\Lambda}\sum_{k=1}^{N-1}b_k
\sum_{l=1}^{k-1}\left ({k\choose l}Z_{l}Z_{k-l}+2Z_k\cos\phi\right )
\\ \nn
-\epsilon^4e^{2\Lambda}\sum_{k=1}^{N-1}b_k
\left (
\sum_{l_1=1}^{k-1}\sum_{l_2=1}^{k-l_1-1}{k\choose l_1}{k-l_1\choose l_2}
Z_{l_1}Z_{l_2}Z_{k-l_1-l_2}
\right .
\\ \label{eq_amp_epsilon}
\left .+3\sum_{l=1}^{k-1}{k\choose l}Z_{l}Z_{k-l}\cos\phi
+3Z_k\cos^2\phi
\right )+{\cal O}(\epsilon^5).
\ee
As before, we can multiply this expression by $\sin\phi$ and $\cos\phi$, and use the relation
\eref{eq_const_2} to obtain independently an equation for $\aamp'$ and $\beta'$ respectively. A time average over the period $2\pi/\Omega_0$ is then performed, after we assumed that $\aamp'$ and $\beta'$ are quasi constant quantities. After some algebra, we obtain the set of ODEs
\bb\nn
\aamp'(\tau)\sum_{k=1}^{N-1}c_k(a_k+\epsilon^2b_k)\simeq
-\frac{3\epsilon^4}{8}e^{2\Lambda}\sum_{k=1}^{N/2}(-1)^{k}b_{2k-1}\Omega_0^{2k-1}
\\ \label{eq_corr}
\beta'(\tau)\sum_{k=1}^{N-1}c_k(a_k+\epsilon^2b_k)\simeq
-\frac{3\epsilon^4}{8}e^{2\Lambda}\sum_{k=1}^{(N-1)/2}(-1)^{k}b_{2k}\Omega_0^{2k},
\ee
where we have dropped the dependence in $\Lambda_0\ll\Omega_0$ in the coefficients in order to
simplify the expressions. We can rewrite the first equation using the substitution $u(\tau)=e^{\Lambda}$, and obtain the relation $u'\simeq\Lambda_0u-3\epsilon^3a_0u^3/8$, with the constant $a_0$ given by
\bb
a_0=\frac{\sum_{k=1}^{N/2}(-1)^{k}b_{2k-1}\Omega_0^{2k-1}}
{\sum_{k=1}^{N-1}c_k(a_k+\epsilon^2b_k)}.
\ee
%
$u$ reaches an asymptotic non-zero limit if $a_0>0$, given by
$u_0^2=8\Lambda_0/(3a_0\epsilon^4)$. If $\Lambda_0<0$, $u_0=0$ and no oscillation is observed. If $\Lambda_0=0$ and $a_0<0$, additional corrections are necessary in order to verify the existence of oscillations. The equation for the frequency shift can be written as $\beta'\simeq-3\epsilon^4
u^2\beta_0/8$, with constant $\beta_0$ given by
\bb
\beta_0=\frac{\sum_{k=1}^{(N-1)/2}(-1)^{k}b_{2k}\Omega_0^{2k}}{\sum_{k=1}^{N-1}c_k(a_k+\epsilon^2b_k)}.
\ee
The shift in frequency will therefore be given asymptotically by $\beta'=-\Lambda_0\beta_0/a_0$,
independent of $\epsilon$. The sign of the shift depends on the sign of $\beta_0$, in the case
where $a_0>0$. We can extend the expansion \eref{eq_amp_epsilon} up to the $\epsilon^6$ order, and find that $\beta'\simeq-\epsilon^4u^2\beta_0(3/8+5\epsilon^2u^2/16)$\footnote{The numerical coefficients $3/8$ and $5/16$ are the angle average of $\cos^4\phi$ and $\cos^6\phi$ respectively.}, which should give a negligible correction. For example, for the parameters given in figure \ref{fig7_eigen_vdiff}
and $N=80$, we obtain numerically $\epsilon\simeq 0.319$, which is small enough to apply the previous expansion. In particular, after solving for the roots of the polynom $R$, we find $z_{max}^+=\Lambda_0+i\Omega_0\simeq 0.026+1.669i$, and therefore $\Lambda_0/\Omega_0\ll 1$. The frequency $\Omega$ from the numerical resolution of the ODEs \eref{eq_gen_vk} was evaluated by Fourier analysis and found to be equal to $\Omega\simeq 1.562$, while the correction $\Omega_0+\beta'\simeq 1.517$ gives a closer value to the numerical estimate than $\Omega_0$. In this case we find that $a_0>0$ and $\beta_0>0$. The amplitude of the oscillations, see figure \ref{fig7_eigen_vdiff}(b), corresponding to $\sqrt{\mu}u_0$, was found after a non-linear curve fitting with a simple cosine function to be equal approximately to $0.071$ while the amplitude given by the expansion \eref{eq_corr} gives $0.047$, for $\vl\simeq 0.044$, which is of the same order of magnitude. The expansion \eref{eq_corr} gives a first estimate of the oscillation characteristics, and shows that these oscillations can be stabilized by
considering the first non-linear terms of the ODE \eref{eq_diff_v1_ab}.

%
\section{Conclusion}
In this paper, we have considered a simple model of coagulation-fragmentation of particle clusters
which displays collective oscillations for a distribution of masses broad enough. The mass-dependent kernel leads to a coefficient matrix that has at least two conjugate complex eigenvalues with strictly positive real part, provided the size of the matrix is sufficiently large. This is a necessary condition for the emergence of a self-oscillating mechanism. This model allows us to evaluate all the densities from the monomer density solution $\vk_1$ that satisfies an implicit non-local Volterra integral equation. This equation depends on two kernels or excitation functions (which are identical in the large $N$ limit) that have a strongly exponential divergence with positive and negative values. However $\vk_1$ appears to oscillate with a finite amplitude and with a minimum positive value proportional to the current. The oscillation amplitude depends on the parameter values such as current, the exponent of the coagulation kernel and the fragmentation rate. When we approximate the kernels with a linear or sawtooth function, we have shown that $\vk_1$ can never be negative, and is larger than $\mu$ when $\mu$ is small in the first approximation.

When we consider a harmonic function as an approximation of the kernels in the long time limit, i.e. with a diverging oscillatory behavior, the model shows self-excited oscillations that can be explained by the presence of a fluctuating negative damping in the expansion of the corresponding
ODE around the threshold beyond which oscillations appear. This phenomenon is very similar to the behavior of a van der Pol oscillator in non-conservative systems which has a quadratic contribution in the damping factor.
When the amplitude increases, the damping factor tends to be positive because of the minus sign in front of the quadratic term, and therefore the amplitude is reduced. And inversely, when the amplitude is small, the damping factor is negative and tends to increase the overall amplitude. The non-conservative
oscillator can sustain self-oscillations only in the presence of a current, which drives the
amplitude. The period of these oscillations does not depend on the current, however no exact solution can be found for its dependence on the other parameters, except around the oscillation threshold using a perturbative calculus, where it is found to be equal in this approximate model to $\sqrt{2}$ times the modulus of the largest eigenvalue of the coefficient matrix. The generalization to $N$ oscillators
leads to a more complex ODE that can be solved using the same perturbative method for some range of small parameters, but it is also the presence of negative damping factors that gives rise to self-oscillations.

Several questions remain concerning the presence of self-oscillations observed in more general aggregation kernels and without source \cite{Brilliantov:2018}. In the present manuscript, the oscillations are driven by a source of monomers, but a more comprehensive study is needed for closed and conservative systems yielding oscillations without any external source, including extended fragmentation processes between clusters of different masses.

\ack
I would like to thank the Department of Physics and Astronomy at the Seoul National University where part of this work was carried out for its hospitality.

\section*{References}

\bibliography{biblio-cluster}
\end{document}